\documentclass{LT23auth}
\usepackage{graphicx}

\begin{document}

\begin{frontmatter}

\title{Flow equation renormalization of a spin-boson model with a structured bath}

\author[address1]{Silvia Kleff\thanksref{thank1}},
\author[address2]{Stefan Kehrein},
\author[address1]{Jan von Delft}

\address[address1]{Lehrstuhl f{\"u}r Theoretische Festk{\"o}rperphysik, Ludwig-Maximilians Universit{\"a}t, 
                   Theresienstr.37, 80333 M\"unchen, Germany}

\address[address2]{Theoretische Physik III -- Elektronische Korrelationen und 
Magnetismus, Universit{\"a}t Augsburg, 86135 Augsburg, Germany}

\thanks[thank1]{E-mail: kleff@theorie.physik.uni-muenchen.de}

\begin{abstract}
We  discuss the dynamics of a spin coupled to a 
damped harmonic oscillator. This system can be mapped 
to a spin-boson model with a structured bath, i.e. the 
spectral function of the bath has a resonance peak. 
We diagonalize the model by means of infinitesimal 
unitary transformations ({\em flow equations}), 
thereby decoupling the small quantum 
system from its environment,
and calculate  spin-spin correlation functions.
\end{abstract}

\end{frontmatter}

\section{Introduction - Model}
\begin{figure}[b]
\begin{center}\leavevmode
\includegraphics[width=0.8\linewidth]{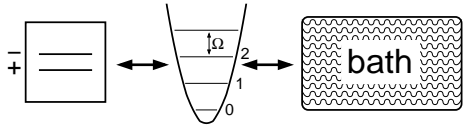}
\caption{A two-level system is coupled to a damped harmonic 
oscillator with frequency $\Omega$.}
\label{fig:system}\end{center}\end{figure}

Recently a new strategy for performing measurements on 
solid state (Josephson) qubits was proposed which uses
the entanglement of the qubit with states of a damped oscillator~\cite{wilhelm}, with this oscillator representing the plasma 
resonance of the Josephson junction. 
 This system of a 
spin coupled to a damped harmonic oscillator 
(see Fig.\ \ref{fig:system}) can be mapped to a standard model 
for dissipative quantum systems, namely the spin-boson model~\cite{garg}.
Here the spectral function governing the dynamics of the 
spin has a resonance peak. Such structured baths were also 
discussed in connection with electron transfer processes~\cite{garg}.
We  use the flow equation method introduced by Wegner~\cite{wegner} 
 to analyze the  system shown in Fig.\ \ref{fig:system}, 
consisting of a two-level system coupled to a harmonic oscillator $\Omega$, 
which is coupled to a bath of harmonic oscillators: 
   \begin{eqnarray}
       \tilde{{\mathcal H}} &=& -\frac{\Delta_0}{2}\sigma_{\mathrm{x}} +\Omega B^{\dagger}B + g(B^{\dagger}+B)\sigma_{\mathrm{z}} 
                           + \sum_{\mathrm{k}}\tilde{\omega}_{\mathrm{k}}\tilde{b}_{\mathrm{k}}^{\dagger}\tilde{b}_{\mathrm{k}}\nonumber\\                         & + &(B^{\dagger}+B)\sum_{\mathrm{k}}\kappa_{\mathrm{k}}(\tilde{b}_{\mathrm{k}}^{\dagger}+\tilde{b}_{\mathrm{k}})
                           + (B^{\dagger}+B)^2\sum_{\mathrm{k}}\frac{\kappa_{\mathrm{k}}^{2}}{\tilde{\omega}_{\mathrm{k}}},\nonumber
   \end{eqnarray}
with  the spectral function 
$J(\omega)\equiv\sum_{\mathrm{k}}\kappa_{\mathrm{k}}^{2}\delta(\omega-\tilde{\omega}_{\mathrm{k}})=\Gamma\omega$. 
This system can be mapped to a spin-boson model~\cite{garg}
   \begin{equation} 
       {\mathcal H}=-\frac{\Delta_0}{2}\sigma_{\mathrm{x}} 
                    +\frac{1}{2}\sigma_{\mathrm{z}}\sum_{\mathrm{k}}\lambda_{\mathrm{k}} (b_{\mathrm{k}}^{\dagger}+b_{\mathrm{k}})
                    +\sum_{\mathrm{k}}\omega_{\mathrm{k}}b_{\mathrm{k}}^{\dagger}b_{\mathrm{k}},\nonumber
   \label{eq:spin_boson}
   \end{equation}
where the dynamics of the spin depends only on the  spectral function
$J(\omega)\equiv\sum_{\mathrm{k}}\lambda_{\mathrm{k}}^{2}\delta(\omega-\omega_{\mathrm{k}})$ 
given by
   \begin{equation} 
       J(\omega)=\frac{2\alpha\omega\Omega^4}{(\Omega^2-\omega^2)^2+(2\pi \Gamma\omega\Omega)^2}
                 \textrm{ with }\alpha=\frac{8\Gamma g^2}{\Omega^2}.
   \end{equation}

\section{Method - Results}
Using the flow equation  technique we approximately diagonalize the   
Hamiltonian ${\mathcal H}$ [Eq.(\ref{eq:spin_boson})] by means of infinitesimal 
unitary transformations. The continuous sequence of unitary 
transformations $U(l)$ is labelled by a flow parameter $l$. 
Applying such a transformation to a given Hamiltonian, this 
Hamiltonian becomes a function of $l$: 
${\mathcal H}(l)=U(l){\mathcal H}U^{\dagger}(l)$.
Here ${\mathcal H}(l=0)={\mathcal H}$  is the initial Hamiltonian 
and ${\mathcal H}(l=\infty)$ is the final diagonal Hamiltonian. 
Usually it is more convenient to work with a differential formulation 
  \begin{equation}
   \label{eq:differentiell}
  \frac{d{\mathcal H}(l)}{dl}=[\eta(l),{\mathcal H}(l)]\quad\textrm{with}\quad
\eta(l)=\frac{dU(l)}{dl}U^{-1}(l). 
  \end{equation}
Using the flow equation approach one can decouple system and 
bath by diagonalizing ${\mathcal H}(l=0)$~\cite{kehrein}:
  \begin{equation}
        {\mathcal H}(l=\infty)=
                      -\frac{\Delta_{\infty}}{2}\sigma_{\mathrm{x}} 
                      + \sum_{\mathrm{k}}\omega_{\mathrm{k}}b_{\mathrm{k}}^{\dagger}b_{\mathrm{k}}.
\label{eq:effective_H}
  \end{equation}
Here $\Delta_{\infty}$ is the renormalized tunneling frequency. 
For the generator of the flow we choose the  Ansatz~\cite{kehrein}:
 \begin{eqnarray}
\eta &=&\sum_{\mathrm{k}}\left(\mathrm{i}\sigma_{\mathrm{y}}\Delta(b_{\mathrm{k}}+b_{\mathrm{k}}^{\dagger})
      +\sigma_{\mathrm{z}}\omega_{\mathrm{k}}(b_{\mathrm{k}}-b^{\dagger}_{\mathrm{k}})\right)\frac{\lambda_{\mathrm{k}}}{2}
        \left(\frac{\Delta-\omega_{\mathrm{k}}}{\Delta+\omega_{\mathrm{k}}}\right)\nonumber\\
   & + &\frac{\Delta}{2}\sum_{\mathrm{q,k}}\lambda_{\mathrm{k}}\lambda_{\mathrm{q}}I(\omega_{\mathrm{k}},\omega_{\mathrm{q}},l)
           (b_{\mathrm{k}}+b_{\mathrm{k}}^{\dagger})(b_{\mathrm{q}}-b_{\mathrm{q}}^{\dagger}),    
 \end{eqnarray}
\[
       \mathrm{with}\quad
I(\omega_{\mathrm{k}},\omega_{\mathrm{q}},l)=
\frac{\omega_{\mathrm{q}}}{\omega_{\mathrm{k}}^{2}-\omega_{\mathrm{q}}^{2}}
\left(\frac{\omega_{\mathrm{k}} -\Delta}{\omega_{\mathrm{k}} +\Delta} 
 + \frac{\omega_{\mathrm{q}}-\Delta}{\omega_{\mathrm{q}}+\Delta}\right).\nonumber
\]
The flow equations for the effective Hamiltonian [Eq.~(\ref{eq:effective_H})] then 
take the following form:
   \begin{eqnarray}
      \label{eq:flussgleichungen} 
      \frac{\partial J(\omega,l)}{\partial l}&=&-2(\omega -\Delta )^2 J(\omega,l)\\\nonumber
      & +& 2\Delta J(\omega,l)\int\mathrm{d}\omega^{\prime}J(\omega^{\prime},l)I(\omega,\omega^{\prime},l),\\
        \frac{d\Delta}{dl} &=& -\Delta\int\mathrm{d}\omega J(\omega,l)\frac{\omega -\Delta}{\omega +\Delta}.
   \end{eqnarray}
The unitary flow diagonalizing the Hamiltonian generates a flow 
for $\sigma_{\mathrm{z}}(l)$ which takes the structure
  \begin{equation}
      \sigma_{\mathrm{z}}(l)=h(l)\sigma_{\mathrm{z}} +
      \sigma_{\mathrm{x}} \sum_{\mathrm{k}}\chi_{\mathrm{k}}(l)(b_{\mathrm{k}}+b_{\mathrm{k}}^{\dagger}),
  \label{eq:sigma_z}
  \end{equation}
where  $h(l)$ and $\chi_{\mathrm{k}}(l)$  obey the 
differential equations
  \begin{eqnarray}
   \frac{dh}{dl} &=& -\Delta\sum_{\mathrm{k}}\lambda_{\mathrm{k}}\chi_{\mathrm{k}}\frac{\omega_{\mathrm{k}}-\Delta}{\omega_{\mathrm{k}}+\Delta},\\
   \frac{d\chi_{\mathrm{k}}}{dl}&=&\Delta h\lambda_{\mathrm{k}}\frac{\omega_{\mathrm{k}}-\Delta}{\omega_{\mathrm{k}}+\Delta}
    +\sum_{\mathrm{q}}\chi_{\mathrm{q}}\lambda_{\mathrm{k}}\lambda_{\mathrm{q}}\Delta I(\omega_{\mathrm{k}},\omega_{\mathrm{q}},l).
\end{eqnarray}
One can show that the function  $h(l)$ decays to zero as $l\rightarrow\infty$. 
Therefore the observable $\sigma_{\mathrm{z}}$ decays completely into bath operators~\cite{kehrein}.

\begin{figure}[t]
\begin{center}\leavevmode
\includegraphics[width=0.99\linewidth]{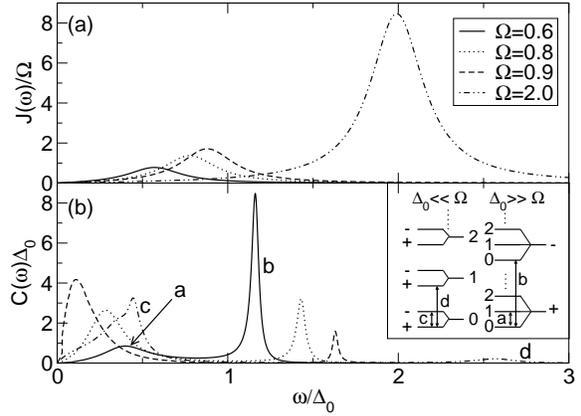}
\caption{(a) Different effective spectral functions $J(\omega,l=0)$ and 
(b) the corresponding $C(\omega)$ for $\Omega\Gamma=0.06$ and $\alpha =0.15$.
The inset shows the term scheme of a two-level system coupled to a harmonic oscillator for the 
two limits $\Delta_0\ll\Omega$ and $\Delta_0\gg\Omega$.}
\label{fig:J+C}\end{center}\end{figure}

We integrated the flow equations numerically in order to calculate the Fourier
transform, $C(\omega)$, of the spin-spin correlation function 
  \begin{equation}
      C(t)\equiv \frac{1}{2}\langle\sigma_{\mathrm{z}}(t)\sigma_{\mathrm{z}}(0)+
        \sigma_{\mathrm{z}}(0)\sigma_{\mathrm{z}}(t)\rangle .
  \end{equation}
$C(t)$ can be used to calculate dephasing and relaxation times for measurements
on qubits~\cite{wilhelm}.  
Fig.~\ref{fig:J+C}(a) shows $J(\omega,l=0)$ and Fig.~\ref{fig:J+C}(b)  
$C(\omega)$ for different values of $\Omega$. $C(\omega)$ displays 
a double-peak structure, which can be understood from the term scheme 
shown in the inset. The  arrows indicate the transitions responsible for  the peaks in 
$C(\omega )$. Additional structure of $C(\omega )$ due to higher order 
transitions in the term scheme is not seen in Fig.~\ref{fig:J+C}. 
This is due to our Ansatz for $\sigma_{\mathrm{z}}(l)$ [see Eq.(\ref{eq:sigma_z})], 
which does not include the corresponding higher order terms. However, we do 
not expect the additional peaks to have much weight, as the sum rule~\cite{kehrein}
for the total spectral weight is fulfilled with an error of less than $5\%$ for
all the plots in
Fig.~\ref{fig:J+C}(b).
 We leave  the extension 
of the Ansatz for $\sigma_{\mathrm{z}}(l)$ for future work.

\begin{ack}
The authors would like to thank F. Wilhelm for helpful discussions.
S.~Kehrein acknowledges support by the SFB~484 of the Deutsche Forschungsgemeinschaft.
\end{ack}

\end{document}